\documentclass[12pt]{article}

\usepackage{amsmath,amssymb,amsthm,mathtools,bm}
\usepackage{hyperref}
\theoremstyle{plain}

\theoremstyle{definition}

\theoremstyle{proposition}

\theoremstyle{lemma}

\theoremstyle{remark}

\usepackage{color}

\usepackage{cite}

\usepackage[pdftex]{graphicx}

\makeatletter
 
  \@addtoreset{equation}{section}
 \makeatother

\newcommand{\nn}{\nonumber}
\newcommand{\pd}{\partial}

\newcommand{\bea}{\begin{eqnarray}}
\newcommand{\eea}{\end{eqnarray}}

\def\X5sp{{\rm X}_5}
\def\Y3sp{{\rm Y}_3}
\def\Z3sp{{\rm Z}_3}

\begin{document}
\setlength{\oddsidemargin}{0cm}
\setlength{\baselineskip}{7mm}

\begin{titlepage}
\begin{flushright}   \end{flushright} 

~~\\

\vspace*{0cm}
    \begin{Large}
       \begin{center}
         {Fundamental structure of string geometry theory}
       \end{center}
    \end{Large}
\vspace{1cm}

\begin{center}
        Matsuo S{\sc ato}$^{*}$\footnote
           {e-mail address : msato@hirosaki-u.ac.jp}
\\
      \vspace{1cm}
       
         {$^{*}$\it Graduate School of Science and Technology, Hirosaki University\\ 
 Bunkyo-cho 3, Hirosaki, Aomori 036-8561, Japan}\\

%
                    
\end{center}

\hspace{5cm}

\begin{abstract}
\noindent
String geometry theory is one of the candidates of a non-perturbative formulation of string theory. In this theory, the ``classical'' action is almost uniquely determined by T-symmetry, which is a generalization of the T-duality, where  the parameter of ``quantum'' corrections $\beta$ in the path-integral of the theory is independent of that of quantum corrections $\hbar$ in the perturbative string theories. We distinguish the effects of $\beta$ and $\hbar$ by putting " " like "classical" and "loops" for tree level and loop corrections with respect to $\beta$, respectively, whereas by putting nothing like classical and loops for tree level and loop corrections with respect to $\hbar$, respectively. A non-renormalization theorem states that there is no ``loop'' correction. Thus, there is no problem of non-renormalizability, although the theory is defined by the path-integral over the fields including a metric on string geometry. 
No ``loop'' correction is also the reason why the complete path-integrals of the all-order perturbative strings in general string backgrounds are derived from the ``tree''-level two-point correlation functions in the perturbative vacua, although string geometry includes information of genera of the world-sheets of the stings. 
Furthermore, a non-perturbative correction in string coupling with the order $e^{-1/g_s^2}$ is given by a transition amplitude representing a tunneling process between the semi-stable vacua  in the ``classical'' potential by an ``instanton'' in the theory.
From this effect, a generic initial state will reach the minimum of the potential.

\end{abstract}

\vfill
\end{titlepage}
\vfil\eject

\setcounter{footnote}{0}

\section{Introduction}\label{intro}
\setcounter{equation}{0}
String geometry theory is one of the candidates of the non-perturbative formulation of string theory. In this theory, the general backgrounds representing pertubative vacua in string theory are identified by the sequential papers \cite{Sato:2017qhj, Sato:2019cno, Sato:2020szq, Sato:2022owj, Sato:2022brv, Sato:2023lls, Nagasaki:2023fnz, futurehetero, Kudo, Honda:2020sbl, Honda:2021rcd}. The path-integrals of all order perturbative strings in general string backgrounds are derived from the ``tree'' level two-point correlation functions of the path-integral of string geometry theory with these perturbative vacua fixed \cite{Sato:2017qhj, Sato:2019cno, Sato:2020szq, Sato:2022owj, Sato:2022brv, Sato:2023lls, Nagasaki:2023fnz, futurehetero, Kudo}, where  the parameter of ``quantum'' corrections $\beta$ in the path-integral of the theory is independent of that of quantum corrections $\hbar$ in the perturbative string theories. We distinguish the effects of $\beta$ and $\hbar$ by putting " " like "classical" and "loops" for tree level and loop corrections with respect to $\beta$, respectively, whereas by putting nothing like classical and loops for tree level and loop corrections with respect to $\hbar$, respectively.
Potentials for string backgrounds are obtained by restricting the ``classical'' potential in string geometry theory to the perturbative vacua  \cite{ Nagasaki:2023fnz, futurehetero, Kudo}. It is conjectured in  \cite{ Nagasaki:2023fnz}  that these potentials represent the string theory landscape and the minimum of them represents the true vacuum in string theory. 

In this paper, we establish a theoretical basis for string geometry theory by solving the following problems arising from the above results. First, is not string geometry theory  non-renormalizable, because the theory is defined by a path-integral over the fields including a metric on string geometry? Why are the complete path-integrals of the all-order perturbative strings  derived from the ``tree''-level two-point correlation functions, although string geometry includes information of genera of the world-sheets of the stings?  Are not ``quantum'' corrections to the ``classical'' potential necessary? On the other hand, how are dynamics of transitions between semi-stable vacua described in string geometry theory, although such  transitions do not occur in a classical theory. Moreover, how is the classical potential determined? 

The organization of the paper is as follows. In section 2, we review string geometry. In subsection 3.1, we define the theory: we regularize the dimensions of the theory  in 3.1.1, define the fields on string geometry in 3.1.2, determine the action in 3.1.3, and give the path-integral in 3.1.4. In subsection 3.2, we describe non-perturbative dynamics. In subsection 3.3, we make a perturbation. In subsection 3.4, we prove a non-renormalization theorem. In subsection 3.5, we discuss the conjecture that the true vacuum will be determined by the minimum of the classical potential restricted to the perturbative vacua. 
In subsection 3.6, the time evolution in string geometry theory is discussed. In section 4,  we conclude and discuss our results.

\vspace{1cm}
\section{String geometry}

String manifold is constructed by patching open sets in string model space $E= \cup_{T} E_T$, where 
$T$ runs  IIA, IIB, I, SO(32) het, and $E_8 \times E_8$ het.  
We call each open set $U \subset E_T$ IIA, IIB, I,  $SO(32)$ heterotic, and $E_8 \times E_8$ heterotic charts, respectively. We will define the coordinates  representing the points composing the model spaces  in the following.

\subsection{String manifolds}

\begin{figure}[htb]
\centering
\includegraphics[width=3cm]{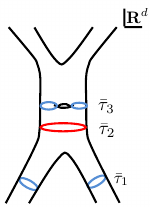}
\caption{Various string states. The red and blue lines represent one string and two strings, respectively.}
\label{states}
\end{figure}

One of the coordinates of the model space is spanned by string geometry time $\bar{\tau} \in \bold{R}$ and another is spanned by super vierbeins $\bar{\bold{E}}_T$, one-to-one corresponding to super Riemann surfaces with punctures $\bar{\bold{\Sigma}}_T \in \mathcal{M}_{T}$ \cite{NotesOnSupermanifolds, WittenSupermoduli, SuperPeriod}. 
On each  $\bar{\bold{\Sigma}}_T$,  
 a global time is defined canonically and uniquely by the real part of the integral of an Abelian differential \cite{Krichever:1987a, Krichever:1987b, Sato:2017qhj}.
We identify this global time as $\bar{\tau}$ and restrict $\bar{\bold{\Sigma}}_T$ to a $\bar{\tau}$ constant hyper surface, and obtain $\bar{\bold{\Sigma}}_T|_{\bar{\tau}}$. 
An embedding of $\bar{\bold{\Sigma}}_T|_{\bar{\tau}}$ to $\mathbb{R}^{d}$  is parametrized by the other coordinates $ \bold{X}_T(\bar{\tau})$.
$\bar{\bold{\Sigma}}_T$ is a union of supercylinders and superstrips at $\bar{\tau}\cong \pm \infty$. Thus, we define a point $ [\bar{\tau},\bar{\bold{E}}_T, \bold{X}_T(\bar{\tau})] $ in the model space as  equivalence classes $ [\bar{\tau}\cong \pm \infty,\bar{\bold{E}}_T, \bold{X}_T(\bar{\tau}\cong \pm \infty)] $ 
 by a relation 
 $(\bar{\tau}\cong \pm \infty,\bar{\bold{E}}_T, \bold{X}_T(\bar{\tau}\cong \pm \infty)) 
 \sim
(\bar{\tau}\cong \pm \infty,\bar{\bold{E}}'_T, \bold{X}'_T(\bar{\tau}\cong \pm \infty))  $ 
if 
 the supercylinders and superstrips are the same at $\bar{\tau}\cong \pm \infty$ and 
 $\bold{X}_T(\bar{\tau}\cong \pm \infty)
 =
 \bold{X}'_T(\bar{\tau}\cong \pm \infty)$,
 and equivalence classes 
 $ [\bar{\tau}\ncong \pm \infty,\bar{\bold{E}}_T, \bold{X}_T(\bar{\tau}\ncong \pm \infty)] $ 
 by a trivial relation 
 $(\bar{\tau}\ncong \pm \infty,\bar{\bold{E}}_T, \bold{X}_T(\bar{\tau}\ncong \pm \infty)) 
 \sim
 (\bar{\tau}\ncong \pm \infty,\bar{\bold{E}}_T, \bold{X}_T(\bar{\tau}\ncong \pm \infty)) $.
Because the bosonic part of $\bar{\bold{\Sigma}}_T|_{\bar{\tau}}$ is isomorphic to $ S^1 \cup \cdots \cup S^1 \cup I^1 \cup \cdots \cup I^1$, where $I^1$
represents a line segment, and $\bold{X}_T (\bar{\tau}): \bar{\bold{\Sigma}}_T|_{\bar{\tau}} \to \mathbb{R}^{d}$, $ [\bar{\tau},\bar{\bold{E}}_T, \bold{X}_T(\bar{\tau})] $ represent many-body strings in $\mathbb{R}^{d}$ as in Fig. \ref{states}.
A sector of the model space $E_T$  is defined by the collection of  $ [\bar{\tau},\bar{\bold{E}}_T, \bold{X}_T(\bar{\tau})] $ by considering all the values of $\bar{\tau}$, all the  $\bar{\bold{E}}_T$, and all the $\bold{X}_T(\bar{\tau})$: 
$E_T= \{ [\bar{\tau},\bar{\bold{E}}_T, \bold{X}_T(\bar{\tau})] \}$.

Here, we will define topologies of $E_T$,  where $E= \cup_{T} E_T$ is a disjoint union. We define an $\epsilon$-open neighborhood of 
$[\bar{\tau}_s,\bar{\bold{E}}_{T}, \bold{X}_{s T} (\bar{\tau}_s)]$
by
\begin{eqnarray}
&&U([\bar{\tau}_s,\bar{\bold{E}}_{T}, \bold{X}_{s T} (\bar{\tau}_s)], \epsilon) \nonumber \\
&:=&
\biggl\{ [\bar{\tau},\bar{\bold{E}}_T, \bold{X}_T(\bar{\tau})]
\bigm| \sqrt{|\bar{\tau}-\bar{\tau}_s|^2
+\| \bold{X}_{T}(\bar{\tau}) -\bold{X}_{s T}(\bar{\tau}_s) \|^2}
<\epsilon  \biggr\}. \label{HeteroNeighbour}
\end{eqnarray}
$U([\bar{\tau}_s\simeq \pm \infty,\bar{\bold{E}}_{T}, \bold{X}_{s T} (\bar{\tau}_s\simeq \pm \infty)], \epsilon) 
=
U([\bar{\tau}_s\simeq \pm \infty,\bar{\bold{E}}'_{T}, \bold{X}'_{s T} (\bar{\tau}_s\simeq \pm \infty)], \epsilon)$
 consistently if the supercylinders and superstrips are the same at $\bar{\tau}\cong \pm \infty$,
 $\bold{X}_T(\bar{\tau}\cong \pm \infty)
 =
 \bold{X}'_T(\bar{\tau}\cong \pm \infty)$, and $\epsilon$ is small enough, because the $\bar{\tau}_s\simeq \pm \infty$ constant hypersurfaces traverses only supercylinders and superstrips  overlapped by $\bar{{\boldsymbol \Sigma}}_T$ and $\bar{{\boldsymbol \Sigma}}'_T$. 
$U$ is defined to be an open set of $E_T$ if there exists $\epsilon$ such that $U([\bar{\tau}_s,\bar{\bold{E}}_{T}, \bold{X}_{s T} (\bar{\tau}_s)], \epsilon)  \subset U$ for an arbitrary point $[\bar{\tau}_s,\bar{\bold{E}}_{T}, \bold{X}_{s T} (\bar{\tau}_s)] \in U$. The topology of $E_T$ satisfies the axiom of topology because the $\epsilon$-open neighborhood is defined by the distance.

By this definition, arbitrary two string states on a connected super Riemann surface in $E_T$ are connected continuously. 
Thus, there is a one-to-one correspondence between a super Riemann surface in $\mathbb{R}^{d}$ and a curve  parametrized by $\bar{\tau}$ from $\bar{\tau}=-\infty$ to $\bar{\tau} = \infty$ on $E_T$. 
That is, curves that represent asymptotic processes on $E_T$ reproduce the right moduli space of the super Riemann surfaces in $\mathbb{R}^{d}$. 
Therefore, a string geometry theory possesses all-order information of superstring theory. 
Indeed, the path integral of perturbative superstrings is derived from the string geometry theory as in \cite{Sato:2017qhj, Sato:2019cno, Sato:2020szq, Sato:2022owj, Sato:2022brv, Nagasaki:2023fnz}. 
The consistency of the perturbation theory determines $d=10$ (the critical dimension).

In order to define structures of manifold, let us consider how generally we can define general coordinate transformations between 
 $ [\bar{\tau},\bar{\bold{E}}_T, \bold{X}_T(\bar{\tau})]  \in U \subset E_T$ and 
 $ [\bar{\tau}',\bar{\bold{E}}'_T, \bold{X}'_T(\bar{\tau}')]\in U' \subset E_T$.   
$\bar{\bold{E}}_T$ does not transform to $\bar{\tau}$ and $\bold{X}_{T}(\bar{\tau})$ and vice versa, because $\bar{\tau}$ and $\bold{X}_{T}(\bar{\tau})$ are continuous variables, whereas $\bar{\bold{E}}_T$ is a discrete variable: $\bar{\tau}$ and $\bold{X}_T (\bar{\tau})$ vary continuously, whereas $\bar{\bold{E}}_T$ varies discretely in a trajectory on $E_T$ by definition of the neighborhoods. 
$\bar{\tau}$ does not transform to $\bar{\sigma}$ and $\bar{\theta}$ and vice versa, 
because the superstring states are defined by $\bar{\tau}$ constant hypersurfaces. Under these restrictions, the most general coordinate transformation is given by 
\begin{eqnarray}
&&[\bar{\tau},\bar{\bold{E}}_T(\bar{\sigma}, \bar{\tau}, \bar{\theta}), \bold{X}_T^{\tilde{\bold{I}}}(\bar{\tau})]\nonumber \\
&\mapsto&[\bar{\tau}'(\bar{\tau}, \bold{X}_T(\bar{\tau})),\bar{\bold{E}}'_T(\bar{\sigma}'(\bar{\sigma}, \bar{\theta}), \bar{\tau}'(\bar{\tau}, \bold{X}_T(\bar{\tau})), \bar{\theta}'(\bar{\sigma}, \bar{\theta})), \bold{X'}_T^{ \tilde{\bold{I}}'(\bar{\sigma}, \bar{\theta})}(\bar{\tau}'(\bar{\tau}, \bold{X}_T(\bar{\tau})))(\bar{\tau}, \bold{X}_T(\bar{\tau}))], \nonumber \\
\label{SuperGeneralCoordTrans2}
\end{eqnarray}
where 
$[\bar{\tau},\bar{\bold{E}}_T(\bar{\sigma}, \bar{\tau}, \bar{\theta}), \bold{X}_T^{\tilde{\bold{I}}}(\bar{\tau})]$
are the components of the coordinates 
$[\bar{\tau},\bar{\bold{E}}_T, \bold{X}_T(\bar{\tau})]$,
and $\tilde{\bold{I}}$ include a part of the coordinates on the super Riemann surfaces, $\bar{\sigma}$ and $ \bar{\theta}$.
$\bar{\bold{E}}_T \mapsto \bar{\bold{E}}'_T$ represents a part of the world-sheet superdiffeomorphism transformation, $\bar{\sigma} \mapsto \bar{\sigma}'(\bar{\sigma}, \bar{\theta})$ and  $\bar{\theta} \mapsto \bar{\theta}'(\bar{\sigma}, \bar{\theta})$.
$\bold{X'}_T^{ \tilde{\bold{I}}'}(\bar{\tau}, \bold{X}_T(\bar{\tau}))$ and $\bar{\tau}'(\bar{\tau}, \bold{X}_T(\bar{\tau}))$ are functionals of $\bar{\tau}$ and $\bold{X}_T(\bar{\tau})$. 
Here, we have extended the model space from $E_T=\{[\bar{\tau},\bar{\bold{E}}_T(\bar{\sigma}, \bar{\tau}, \bar{\theta}), \bold{X}_T^{\tilde{\bold{I}}}(\bar{\tau})]\}$ to $E_T=\{[\bar{\tau}',\bold{E}'_T(\bar{\sigma}', \bar{\tau}', \bar{\theta}'), \bold{X'}_T^{\tilde{\bold{I}}'}(\bar{\tau}')]\}$
 by including the points generated by the superdiffeomorphisms $\bar{\sigma} \mapsto \bar{\sigma}'(\bar{\sigma}, \bar{\theta})$,  $\bar{\theta}^{\alpha} \mapsto \bar{\theta}^{'\alpha}(\bar{\sigma}, \bar{\theta})$, and $\bar{\tau} \mapsto \bar{\tau}'(\bar{\tau})$. In the following, $\bar{\sigma}$, $\bar{\tau}$, $\bar{\theta}$ and $\bar{\bold{E}}_T$ represent the coordinates in this extended space.  
We consider all the manifolds which are constructed by patching open sets of the model space $E_T$ by general coordinate transformations (\ref{SuperGeneralCoordTrans2}) and call them string manifolds $\mathcal{M}_T$. 

In the neighborhood,  $\bar{\tau}$ and $ \bold{X}_T(\bar{\tau})$ have the same weights, and we impose diffeomorphism invariance that mixes $\bar{\tau}$ and $ \bold{X}_T(\bar{\tau})$   completely to the theory so that it has the maximal symmetry. Thus, we rename $\bar{\tau}$, $ \bold{X}^d_T$ and rewrite $ [\bar{\tau},\bar{\bold{E}}_T, \bold{X}_T^{\tilde{\bold{I}}}(\bar{\tau})]$ as $ [\bar{\bold{E}}_T, \bold{X}_T^{\bold{I}}(\bar{\tau})]$, where $\bold{I}=d, \tilde{\bold{I}}$. The cotangent space is spanned by $d \bold{X}_T^{\bold{I}}$.
 $d \bar{\bm E}_T$   cannot be a part of basis that span the cotangent space because $ \bar{\bm E}_T$ is a discrete variable as in (\ref{HeteroNeighbour}).  The line elements on string manifolds are defined by inner products of the cotangent vectors,  
 \begin{equation}
 ds^2=
 \bold{G}_{\bold{I}\bold{J}}(\bar{\tau},\bar{\bold{E}}_T, \bold{X}_T(\bar{\tau}))
 d \bold{X}_T^{\bold{I}}d \bold{X}_T^{\bold{J}}.
 \end{equation} 
Here, we should note that the fields are functionals of $\bar{\bm E}_T$. 
The scalar $\bold{\Phi}(\bar{\tau},\bar{\bold{E}}_T, \bold{X}_T(\bar{\tau}))$
and tensors $\bold{B}_{\bold{I}\bold{J}}(\bar{\tau},\bar{\bold{E}}_T, \bold{X}_T(\bar{\tau})), \cdots$ are also defined in the same way because the basis of the cotangent space is given explicitly as $d \bold{X}_T^{\bold{I}}$.

\subsection{Type IIA, IIB, and I charts}

In this subsection, we define type IIA, IIB and I model spaces $E_{T}$ where $T$ runs $IIA$, $IIB$ and $I$. In this space, $\tilde{\bold{I}}= (\mu \bar\sigma  \bar\theta)$, where $\mu=0, 1, \cdots d-1$.
${\boldsymbol X}_{T}^{(\mu \bar\sigma  \bar\theta)}(\bar{\tau})=X^{\mu}(\bar\sigma, \bar\tau)+ \bar{\theta}^{\alpha} \psi^{\mu}_{\alpha}(\bar\sigma, \bar\tau)+\frac{1}{2} \bar{\theta}^2 F^{\mu} (\bar\sigma, \bar\tau)$, where $\psi_{\alpha}^{\mu}$ is a Majorana fermion and  $F^{\mu}$ is an auxiliary field. We abbreviate $T$ and $(\bar{\tau})$ of $X^{\mu}$, $\psi_{\alpha}^{\mu}$ and $F^{\mu}$.

 We define the Hilbert space in these coordinates by the states only with GSO projection, $e^{\pi i F}=1$ and $e^{\pi i \tilde{F}}=(-1)^{\tilde{\alpha}}$ for $T=$ IIA and $e^{\pi i F}=e^{\pi i \tilde{F}}=1$ for $T=$ IIB and I, where $F$ and $\tilde{F}$ are left- and right-handed fermion numbers respectively, and $\tilde{\alpha}$ is 1 or 0 when the right-handed fermion is periodic (R sector) or anti-periodic (NS sector), respectively. $\Omega$ projection is imposed for $T=I$.

The distance in these charts is defined by
\begin{eqnarray}
&&
\| \bold{X}_{T} (\bar{\tau})-\bold{X}_{s \, T}(\bar{\tau}_s) \|^2 \nonumber \\
&:=&\int_0^{2\pi}  d\bar{\sigma} 
\Bigl(|X(\bar{\tau}, \bar{\sigma})-X_s(\bar{\tau}_s, \bar{\sigma})|^2 
+(\bar{\psi}(\bar{\tau}, \bar{\sigma})-\bar{\psi}_s(\bar{\tau}_s, \bar{\sigma}))
(\psi(\bar{\tau}, \bar{\sigma})-\psi_s(\bar{\tau}_s, \bar{\sigma})) \nonumber \\
&+&|F(\bar{\tau}, \bar{\sigma})-F_s(\bar{\tau}_s, \bar{\sigma})|^2 \Bigr).
\end{eqnarray}

Let us define a summation over $\bar{\sigma}$ and $\bar{\theta}$ that is invariant under $(\bar{\sigma}, \bar{\theta}^{\alpha}) \mapsto (\bar{\sigma}'(\bar{\sigma}, \bar{\theta}), \bar{\theta}^{'\alpha}(\bar{\sigma}, \bar{\theta}))$ and transformed as a scalar under $\bar{\tau} \mapsto \bar{\tau}'(\bar{\tau}, \bold{X}_T(\bar{\tau}))$. First, $\int d\bar{\tau} \int d\bar{\sigma}d^2\bar{\theta} \bar{\bold{E}}_T(\bar{\sigma}, \bar{\tau}, \bar{\theta}^{\alpha})$ is invariant under $(\bar{\sigma}, \bar{\tau}, \bar{\theta}^{\alpha}) \mapsto (\bar{\sigma}'(\bar{\sigma}, \bar{\theta}), \bar{\tau}'(\bar{\tau}, \bold{X}_T(\bar{\tau})), \bar{\theta}^{'\alpha}(\bar{\sigma}, \bar{\theta}))$, where $\bar{\bold{E}}_T(\bar{\sigma}, \bar{\tau}, \bar{\theta}^{\alpha})$ is the superdeterminant of $\bar{\bold{E}}_{T M}^{\quad A}(\bar{\sigma}, \bar{\tau}, \bar{\theta}^{\alpha})$. The lapse function, $\bar{n}$ transforms as an one-dimensional vector in the $\bar{\tau}$ direction: 
$\int d\bar{\tau} \bar{n}$ is invariant under $\bar{\tau} \mapsto \bar{\tau}'(\bar{\tau}, \bold{X}_T(\bar{\tau}))$ and transformed as a superscalar under $(\bar{\sigma}, \bar{\theta}^{\alpha}) \mapsto (\bar{\sigma}'(\bar{\sigma}, \bar{\theta}), \bar{\theta}^{'\alpha}(\bar{\sigma}, \bar{\theta}))$. Therefore, 
$\int d\bar{\sigma}d^2\bar{\theta} \hat{\bold{E}}_T(\bar{\sigma}, \bar{\tau}, \bar{\theta}^{\alpha})$,
where
$\hat{\bold{E}}_T(\bar{\sigma}, \bar{\tau}, \bar{\theta}^{\alpha})
:=
\frac{1}{\bar{n}}\bar{\bold{E}}_T(\bar{\sigma}, \bar{\tau}, \bar{\theta}^{\alpha})$,
is transformed as a scalar under $\bar{\tau} \mapsto \bar{\tau}'(\bar{\tau}, \bold{X}_T(\bar{\tau}))$ and invariant under $(\bar{\sigma}, \bar{\theta}^{\alpha}) \mapsto (\bar{\sigma}'(\bar{\sigma}, \bar{\theta}), \bar{\theta}^{'\alpha}(\bar{\sigma}, \bar{\theta}))$.
As a result, the type IIA, IIB and I parts of any action are invariant under $\mathcal N = (1,1)$ supersymmetry transformation because all the indices are contracted by the summation.
These supersymmetries correspond to the world-sheet supersymmetries perturbatively and the target-space supersymmetry is not assumed. When a background is fixed, the theory around it has a target-space supersymmetry only when the background is special.

\subsection{$SO(32)$ and $E_8 \times E_8$ heterotic charts}
In this subsection, we define heterotic model spaces, $E_{G \mbox{het}}$ where $G$ runs $SO(32)$ and $E_8 \times E_8$. In this space, $\tilde{\bold{I}}= (\mu \bar\sigma  \bar\theta), (A \bar\sigma  \bar\theta^-)$, where $\mu=0, 1, \cdots d-1$, $A= 1, \cdots 32$ and 
$\bar\theta^{-}$ has the opposite chirality to $\bar\theta$. 
${\boldsymbol X}_{G}^{(\mu \bar\sigma  \bar\theta)}(\bar{\tau})=X^{\mu}(\bar\sigma, \bar\tau)+ \bar{\theta} \psi^{\mu}(\bar\sigma, \bar\tau) $ and $\bm X_{LG}^{(A \bar\sigma  \bar\theta^-)}(\bar\tau) =  \bar\theta^{-} \lambda_{G}^A(\bar\sigma, \bar\tau)$,  where  $\psi^{\mu}$ and $\lambda_{G}^A$ are Majorana-Weyl fermions with opposite chiralities.
We abbreviate $G$ of $X^{\mu}$ and $\psi^{\mu}$. 
We can define worldsheet fermion numbers of states in a Hilbert space because the states consist of the fields over the local coordinates ${\boldsymbol X}_{G}^{(\mu \bar\sigma  \bar\theta)}(\bar{\tau})=X^{\mu}(\bar\sigma, \bar\tau)+ \bar{\theta} \psi^{\mu}(\bar\sigma, \bar\tau)$ and $\bm X_{LG}^{(A \bar\sigma  \bar\theta^-)}(\bar\tau) =  \bar\theta^{-} \lambda_{G}^A(\bar\sigma, \bar\tau)$. 
For $G=SO(32)$, we take periodicities 
\begin{equation}
\lambda_{SO(32)}^A(\bar{\tau}, \bar{\sigma}+2\pi)=\pm \lambda_{SO(32)}^A(\bar{\tau}, \bar{\sigma}) \quad (A=1, \cdots 32)
\end{equation}
with the same sign on all 32 components. 
We define the Hilbert space in these coordinates by the GSO projection of the states with $e^{\pi i F}=1$ and $e^{\pi i \tilde{F}}=1$, where $F$ and $\tilde{F}$ are the numbers of left- and right- handed fermions $\lambda^A_{SO(32)}$ and $\psi^{\mu}$, respectively. 
For $G=$ $E_8 \times E_8$, the periodicity is given by 
\begin{eqnarray}
\lambda_{E_8 \times E_8}^A(\bar{\tau}, \bar{\sigma}+2\pi)= 
\left\{
\begin{array}{c}
\eta \lambda_{E_8 \times E_8}^A(\bar{\tau}, \bar{\sigma}) \quad (1 \leqq A \leqq 16) \\
\eta' \lambda_{E_8 \times E_8}^A(\bar{\tau}, \bar{\sigma}) \quad (17 \leqq A \leqq 32), 
\end{array}
\right.
\end{eqnarray}
with the same sign $\eta(= \pm 1)$ and $\eta'(= \pm 1)$ on each 16 components.  
The GSO projection is given by $e^{\pi i F_1}=1$, $e^{\pi i F_2}=1$ and $e^{\pi i \tilde{F}}=1$, where $F_1$, $F_2$ and $\tilde{F}$ are the numbers of $\lambda_{E_8 \times E_8}^{A_1}$ ($A_1=1, \cdots, 16$), $\lambda_{E_8 \times E_8}^{A_2}$ ($A_2=17, \cdots, 32$) and $\psi^{\mu}$, respectively.

The distance in these charts is defined by
\begin{eqnarray}
&&
\| \bold{X}_{G}(\bar{\tau}) -\bold{X}_{s G}(\bar{\tau}_s) \|^2 \nonumber \\
&:=&\int_0^{2\pi}  d\bar{\sigma} 
\Bigl(|X(\bar{\tau}, \bar{\sigma})-X_s(\bar{\tau}_s, \bar{\sigma})|^2 
+(\bar{\psi}(\bar{\tau}, \bar{\sigma})-\bar{\psi}_s(\bar{\tau}_s, \bar{\sigma}))
(\psi(\bar{\tau}, \bar{\sigma})-\psi_s(\bar{\tau}_s, \bar{\sigma}))\nonumber \\
&& \qquad \quad +
( \bar{\lambda}_{G}(\bar{\tau}, \bar{\sigma})-\bar{\lambda}_{s G}(\bar{\tau}_s, \bar{\sigma}))
(\lambda_{G}(\bar{\tau}, \bar{\sigma})-\lambda_{s G}(\bar{\tau}_s, \bar{\sigma})  \Big).
\end{eqnarray}

The summations over $(\bar\sigma, \bar\theta)$ and $(\bar\sigma',\bar\theta^{-})$ are defined by 
$\displaystyle\int d\bar\sigma d\bar\theta\hat{\bm E}(\bar\sigma, \bar\tau, \bar\theta)$ and 
$\displaystyle\int d\bar\sigma' d\bar\theta^{-}\bar e(\bar\sigma', \bar\tau)$, respectively. $\hat{\bm E}(\bar{\sigma}, \bar{\tau}, \bar{\theta})\coloneqq 
(1/\bar{n})\bar{\bm E}(\bar{\sigma}, \bar{\tau}, \bar{\theta})$.
These summations are transformed as scalars under $\bar\tau \mapsto \tilde{\bar{\tau}}(\bar\tau, \bm X_{G}(\bar\tau), \bm X_{L G}(\bar\tau))$.  Moreover, $\displaystyle\int d\bar\sigma d\bar\theta\hat{\bm E}(\bar\sigma, \bar\tau, \bar\theta)$ is invariant under a supersymmetry transformation $(\bar{\sigma}, \bar{\theta}) \mapsto (\tilde{\bar{\sigma}}(\bar{\sigma}, \bar{\theta}), \tilde{\bar{\theta}}(\bar{\sigma}, \bar{\theta}))$.
$\displaystyle\int d\bar\sigma' d\bar\theta^{-}\bar e(\bar\sigma', \bar\tau)$ is also invariant under this supersymmetry transformation, because 
$(\mu\bar\sigma\bar\theta)$ and $(A\bar\sigma' \bar\theta^{-})$ in $I = \{d,(\mu\bar\sigma\bar\theta),(A\bar\sigma' \bar\theta^{-})\}$
are independent indices and then $(A\bar\sigma' \bar\theta^{-})$ is not transformed under the supersymmetry.  
As a result, the heterotic part of any action is invariant under this $\mathcal N = (1,0)$ supersymmetry transformation because all the indices are contracted by the summations.

\section{Basis of string geometry theory}

\subsection{Definition of the theory}

\subsubsection{Regularization of dimensions}
The dimension of the space spanned by $\bold{X}_T^{(\mu \bar{\sigma} \bar{\theta})}$ is given by $\sum_{\mu}  \int d\bar{\sigma} d^2 \bar{\theta}\hat{\bar{\bold{E}}}_T \delta^{(\mu \bar{\sigma} \bar{\theta})}_{(\mu \bar{\sigma} \bar{\theta})}$, which is the trace of ``1'' acting on the space explicitly $ \delta^{(\mu \bar{\sigma} \bar{\theta})}_{(\nu \bar{\sigma}' \bar{\theta}')}$. We need a regularization to define the dimension because $\delta^{\bar{\sigma}}_{\bar{\sigma}}=\infty$ and $\delta^{\bar{\theta}}_{\bar{\theta}}=0$. We can regularize $\delta^{\bar{\sigma}}_{\bar{\sigma}'}$
as 
\begin{equation}
\delta^{\,\,\, \bar{\sigma}}_{\epsilon\, \bar{\sigma}'}
=
\frac{1}{\sqrt{\pi} \epsilon} e^{-\frac{(\bar{\sigma}-\bar{\sigma}')^2}{\epsilon^2}}
\end{equation}
and 
$\delta^{\bar{\theta}}_{\bar{\theta}'}$
as
\begin{equation}
\delta^{\,\,\, \bar{\theta}}_{\epsilon \, \bar{\theta}'}
=
(\bar{\theta}-\bar{\theta}')(\bar{\bar{\theta}}-\bar{\bar{\theta}}')
+\epsilon^{\alpha}
(\bar{\theta}+\bar{\theta}')(\bar{\bar{\theta}}+\bar{\bar{\theta}}'),
\label{deltatheta}
\end{equation}
where $\alpha > 0$.
Not only 
$\delta^{\,\,\, \bar{\sigma}}_{\epsilon\, \bar{\sigma}'}$
but also 
$\delta^{\,\,\, \bar{\theta}}_{\epsilon \, \bar{\theta}'}$
has the property of delta function,
\begin{eqnarray}
\lim_{\epsilon \to 0} \int d^2 \bar{\theta}
\delta^{\,\,\, \bar{\theta}}_{\epsilon \, \bar{\theta}'}
f(\bar{\theta})
=
f(\bar{\theta}').
\end{eqnarray}
In addition, 
\begin{equation}
(\delta^{\,\,\, \bar{\theta}}_{\epsilon \, \bar{\theta}'})^2=8\epsilon^{\alpha}\bar{\theta}\bar{\theta}' \bar{\bar{\theta}} \bar{\bar{\theta}}'
\end{equation}
is not zero.  Although 
$(\delta^{\,\,\, \bar{\theta}}_{\epsilon \, \bar{\theta}'})^n=0$  $(n \ge 3)$, 
$(\delta^{\,\,\, \bar{\theta}}_{\epsilon \, \bar{\theta}'})^n$  $(n \ge 3)$ do not appear in the action because all the pair of indices are contracted.  Moreover, a product of
$\delta^{\,\,\, \bar{\sigma}}_{\epsilon\, \bar{\sigma}}$
and
$\delta^{\,\,\, \bar{\theta}}_{\epsilon \, \bar{\theta}}$
is not zero:
\begin{equation}
\delta^{\,\,\, \bar{\sigma}}_{\epsilon\, \bar{\sigma}}
\delta^{\,\,\, \bar{\theta}}_{\epsilon \, \bar{\theta}}
=
\frac{4}{\sqrt{\pi}} 
\theta\bar{\theta}
\epsilon^{\alpha-1}.
\end{equation}
 As a result, the dimension of the space, namely the trace of ``1,''
\begin{equation}
\delta^{\,\,\, (\mu \bar{\sigma} \bar{\theta})}_{\epsilon \, (\nu \bar{\sigma}' \bar{\theta}')}=\frac{1}{\hat{\bar{\bold{E}}}_T}
\delta^{\mu}_{\nu} \delta^{\,\,\, \bar{\sigma}}_{\epsilon\, \bar{\sigma}'}
\delta^{\,\,\, \bar{\theta}}_{\epsilon \, \bar{\theta}'}\end{equation}
is given by 
\begin{equation}
\sum_{\mu}  \int d\bar{\sigma} d^2 \bar{\theta}\hat{\bar{\bold{E}}}_T \delta^{\,\,\, (\mu \bar{\sigma} \bar{\theta})}_{\epsilon (\mu \bar{\sigma} \bar{\theta})}=8\sqrt{\pi}d
\epsilon^{\alpha-1}.
\end{equation}
This is infinity ($\alpha < 1$),  $8\sqrt{\pi}d$ ($\alpha=1$) and zero ($\alpha > 1$). We choose $\alpha < 1$ because the configuration space of strings should be infinite dimensional.  

The flat metric 
$ \bold{G}^{flat}_{(\mu \bar{\sigma} \bar{\theta}) (\nu \bar{\sigma}' \bar{\theta}')}= \delta_{(\mu \bar{\sigma} \bar{\theta}) (\nu \bar{\sigma}' \bar{\theta}')}=\eta_{\mu \nu} \delta(\bar{\sigma}, \bar{\sigma}') (\bar{\theta}-\bar{\theta}')^2$
is invariant under the rigid super translation
$\theta \to \theta +\eta$, 
although ``1'', 
$ \delta^{\,\,\, (\mu \bar{\sigma} \bar{\theta})}_{\epsilon \, (\nu \bar{\sigma}' \bar{\theta}')}$
is not invariant because of the definition (\ref{deltatheta}). The inverse of the flat metric is 
$ \bold{G}_{flat}^{(\mu \bar{\sigma} \bar{\theta}) (\nu \bar{\sigma}' \bar{\theta}')}= \delta^{\,\,\, (\mu \bar{\sigma} \bar{\theta})}_{\epsilon \, (\nu \bar{\sigma}' \bar{\theta}')}$.

\subsubsection{Fields on string manifolds}


Fields on string manifolds are superfields, where we obtain supersymmetric multiplets by expanding them with respect to $\bar{\theta}$ in the indices. Thus, the fields are categorized to scalars, gauge fields, metrics, and completely anti-symmetric tensors.

In a non-perturbative formulation of string theory, it is natural that the backgrounds in the theory include string backgrounds, which parametrize perturbative vacua in string theory, namely the fields in supergravities. Actually, backgrounds in string geometry theory include string backgrounds as 
\begin{eqnarray}
\bar{\bold{G}}_{(\mu\bar{\sigma}\bar{\theta})(\nu\bar{\sigma}'\bar{\theta}')}&=&
\bar{G}_{\mu\nu}(\bold{X}(\bar{\sigma}, \bar{\theta})) \frac{\bar{e}^3}{\sqrt{\bar{h}}}\delta_{\bar{\sigma}\bar{\sigma}'}
\delta_{\bar{\theta} \bar{\theta}'},
\nonumber
\end{eqnarray}
where 
$\bar{G}_{\mu\nu}(x)$ is a background of the ten-dimensional metric. Similarly, backgrounds of the other fields in string geometry theory
$\bold{B}_{\bold{I}\bold{J}}$, 
$\bold{\Phi}$, 
$\bold{A}_{\bold{I}}$  ($N \times N$ matrices), and
$\bold{C}^p_{\bold{I}_1 \cdots \bold{I}_p}$ $(p=0, 1, \cdots )$
include 
backgrounds of NS-NS B field, dilaton, non-Abelian gauge fields, and R-R fields, respectively. 
To extend the fields minimally, we restrict the fields only on regions where the string backgrounds are defined. That is, 
$\bold{A}_{\bold{I}}$ is non-zero only on the points representing open strings in type II chart, and on the points in type I and heterotic charts. 
$\bold{C}^p_{\bold{I}_1 \cdots \bold{I}_p}$
are non-zero only on the points in type I and II charts.
The action of string geometry theory is constructed by these fields because they are the minimum set that has the backgrounds including all the string backgrounds.

\subsubsection{Determine the action}
We impose the following conditions (i), (ii), and (iii) to the action of string geometry theory. 
First, the condition (i) is that the action includes only up to the second derivative terms so as not to have a ghost term\footnote{Generically, a higher derivative term causes a ghost term, whereas there exist exceptions  \cite{Ostrogradsky:1850fid}.}. For example, a propagator
$\frac{1}{(p^2)^2-m^4}=\frac{1}{p^2-m^2} \frac{1}{p^2+m^2}$
has a pole of a ghost mode. 

If a dimensional reduction in one direction is performed, the type IIA and IIB string backgrounds one-to-one correspond to each other under T-dual transformation.  In addition, the effective actions of them, namely dimensionally reduced type IIA and IIB supergravities are T-dual to each other. Similarly, dimensionally reduced fields in string geometry theory  corresponding to the type IIA and IIB string backgrounds one-to-one correspond to each other under generalized T-dual transformation \cite{Sato:2023lls}. In order that string geometry theory is a non-perturbative formulation of string theory, it needs to include all the vacua, namely type IIA, IIB, I, SO(32) heterotic and $E_8 \times E_8$ heterotic vacua, and the action needs to be invariant under the T-dual transformation among these vacua. 
Thus, the condition (ii) is that in the closed string sector of type II charts,  where $\bold{A}_{\bold{I}}=0$, the dimensionally reduced action of sting geometry theory is invariant under the generalized T-duality transformation between type IIA and IIB vacua.  Similarly, the condition (iii) is that in the heterotic charts, where $\bold{C}_{\bold{I}_1 \cdots \bold{I}_p}=0$, it is also invariant under  the transformation between $SO(32)$ and $E_8 \times E_8$ heterotic vacua.  

An action
\begin{eqnarray}
S=\frac{1}{2\beta}\int \mathcal{D}\bold{E}_T \mathcal{D}\bar{\tau} \mathcal{D}\bold{X}_T
\sqrt{-\bold{G}} \left( e^{-2 \bold{\Phi}} \left( \mathbf{R}  + 4 \nabla_{\bold{I}} \bold{\Phi} \nabla^{\bold{I}} \bold{\Phi} - \frac{1}{2} |\tilde{\mathbf{H}} |^{2} -\frac{\alpha'}{4} \mbox{tr}(|\mathbf{F}|^2)  \right) -\frac{1}{2}\sum_{p=1}^{\infty}
|\tilde{\bold{F}}_p|^2     
\right),
\label{action of bos string-geometric model}
\end{eqnarray}
satisfies the above conditions (i), (ii) and (iii): (i) is trivially satisfied. (ii) was shown in \cite{Sato:2023lls}.
The extension to the invariance of (\ref{action of bos string-geometric model}) with $\bold{C}_{\bold{I}_1 \cdots \bold{I}_p}=0$, namely (iii) can be shown in the same way as in (ii) because the dimensionally reduced action of  the heterotic supergravity is invariant under the $SO(32)$ and $E_8 \times E_8$ heterotic T-duality transformation \cite{Bergshoeff:1995cg}. 
This action consists of  a scalar curvature    $\mathbf{R}$ of a metric $\mathbf{G}_{\bold{I}_{1} \bold{I}_{2}}$, a scalar field $\bold{\Phi}$, a field strength $\tilde{\mathbf{H}}=d\mathbf{B}-\alpha'\boldsymbol{\omega}_3$ of a two-form field $\mathbf{B}_{\bold{I}_{1} \bold{I}_{2}}$, where
$\boldsymbol{\omega}_3=\mbox{tr}(\mathbf{A}\wedge d\mathbf{A} -\frac{2i}{3}\mathbf{A} \wedge \mathbf{A} \wedge \mathbf{A})$, and 
$\mathbf{A}$ is a $N \times N$ Hermitian gauge field, whose field strength is given by $\mathbf{F}$, and p-forms $\tilde{\bold{F}}_p$. $\tilde{\bold{F}}_p$ are defined by 
 $\tilde{\bold{F}}_p=d\bold{C}_{p-1}+d\bold{B}\wedge\bold{C}_{p-3}.$
 
 The scalars in the action are only $\bold{\Phi}$ and $\bold{C}_{0}$. There is no invariant potential term under diffeomorphism and the generalized T-dual transformation, because $\bold{\Phi}$ and $\bold{C}_{0}$ are transformed to parts of the metric and $\bold{C}_{1}$, respectively under the generalized T-dual transformation. Therefore, an action that satisfies (i), (ii) and (iii) will be almost uniquely determined to  
 (\ref{action of bos string-geometric model}).
 
The coupling constants in the action are only an overall constant and a constant in front of the gauge field as in (\ref{action of bos string-geometric model})  because of T-symmetry. The overall constant is an inverse of a parameter  $\beta$, which gives  ``quantum'' corrections in the path-integral of the string geometry theory, and is independent of the parameter $\hbar$, which gives quantum corrections.  $\hbar$ is the parameter of the loop expansions of perturbative strings, which are included in string geometry even in the tree level of the path-integral of string geometry theory.  The  constant in front of the gauge field is $\alpha'$, which gives dimensions to the string coordinates as $\frac{1}{\sqrt{\alpha'}}\bold{X}_T^{(\mu \bar{\sigma} \bar{\theta})}$.

\subsubsection{The direction of the negative sign of the metric and the path-integral}
 The direction of the negative sign of the metric is defined to be the direction of $x^0$, which is the zero mode of $\bold{X}_T^{(0 \bar{\sigma} \bar{\theta} )}$. All the other directions have the positive sign. Especially, the direction of string geometry time $\bar{\tau}$ has the positive sign.
Thus, string geometry theory is a Lorentzian theory. Then, the path-integral is given by  
 \begin{eqnarray}
Z=\int \mathcal{D}\bold{G} \mathcal{D}\bold{\Phi}\mathcal{D}\bold{B}\mathcal{D}\bold{C}\mathcal{D}\bold{A} e^{\frac{i}{\beta}S},\label{pathint}
\end{eqnarray}
where the action is redefined by extracting the overall constant $1/\beta$.

 \subsection{Non-perturbative dynamics of changing backgrounds}

For simplicity, we write the path-integral (\ref{pathint}) as  
 \begin{eqnarray}
Z(\psi_f, \psi_i)=\int^{\psi_f}_{\psi_i} \mathcal{D}\psi e^{\frac{i}{\beta} S}, \label{simplepathint}
\end{eqnarray}
where $\psi$ represent arbitrary fields ($\bold{G}_{IJ}, \cdots$) and $\psi_i$ and $\psi_f$ are the initial and final values of $\psi$, respectively. 

In general,  a path-integral is a wave function that satisfies Schrodinger equation, 
\begin{eqnarray}
i \beta \frac{\partial}{\partial \bar{\tau}} Z
=H(-i \beta  \frac{\partial}{\partial \psi_f}, \psi_f) Z,
\end{eqnarray}
where $H(\pi, \psi_f)$ is the Hamiltonian and $\pi$ are the conjugate momenta of $\psi_f$.
Because string geometry theory is a gravitational system, the Hamiltonian can be written as
\begin{equation}
H=NH_0+\sum_{i}N^iH_i+ \sum_{\alpha} N^{\alpha}G_{\alpha}, \label{hamiltonian}
\end{equation}
where $N$, $N^i$ and $N^{\alpha}$ are auxiliary  fields. $N$ and $N^i$ are called Lapse function and shift vector, respectively.  By varying the action with respect to these auxiliary fields, we obtain constraints 
$H_0=0$, $H_i=0$ and $G_{\alpha}=0$, which are called Hamiltonian constraint, momentum constraint, and Gauss laws, respectively. Therefore, 
$\frac{\partial}{\partial \bar{\tau}} Z(\psi_f, \psi_i)=0$, that is the path-integral does not depend on string geometry time $\bar{\tau}$, but depends only on our time $x^0$. Thus,  Schrodinger equation becomes Wheeler de Witt equations,
\begin{eqnarray}
&&H_0(-i \beta  \frac{\partial}{\partial \psi_f}, \psi_f) Z(\psi_f, \psi_i)=0  \label{Hamiltonian} \\
&&H_i(-i \beta  \frac{\partial}{\partial \psi_f}, \psi_f) Z(\psi_f, \psi_i)=0 \label{momentum} \\
&&G_{\alpha}(-i \beta  \frac{\partial}{\partial \psi_f}, \psi_f) Z(\psi_f, \psi_i)=0. \label{Gauss}
\end{eqnarray}
For example, see \cite{Sato:2002kv}. 
(\ref{momentum}) and (\ref{Gauss}) are the conditions of diffeomorphism invariance in all the directions of $X^{(\mu \sigma \theta)}$ including $x^0$, and of gauge invariances, respectively. 
 We expand the path-integral semi-''classically'' with respect to $\beta$, which gives ``quantum'' corrections in the path-integral of string geometry theory:
\begin{equation}
Z(\psi_f, \psi_i)=e^{\frac{i}{\beta}(S_0(\psi_f, \psi_i)+ \beta S_1(\psi_f, \psi_i) +  \beta^2 S_2(\psi_f, \psi_i) + \cdots)}.
\end{equation}
For $S_0(\psi_f, \psi_i)$, Wheeler de Wit equation (\ref{Hamiltonian}) reduces to Hamilton Jacobi equation,
\begin{equation}
H_0( \frac{\partial S_0(\psi_f, \psi_i)}{\partial \psi_f}, \psi_f)=0.
\end{equation}
Namely, $S_0(\psi_f, \psi_i)$ is obtained by substituting the solution $\bar{\psi}$ of the equations of motion $\frac{\delta S}{\delta \psi}=0$ with the initial values $\psi_i$ and the final values $\psi_f$, to the original action $S$. For example, see \cite{Sato:2003ky}. 
The solution  $\bar{\psi}$  can be complex-valued functions in general. Because $\psi$ are real fields,  the solution $\bar{\psi}$ are a ``classical'' solution if $\bar{\psi}$ are real valued, whereas $\bar{\psi}$ are an instanton solution passing through a ``classically'' forbidden region
 if $\bar{\psi}$ are complex valued. The dependence on the constant expectation value of dilaton $\bar{\Phi}_0$  is given by $S_0(\psi_f, \psi_i)=e^{-2 \bar{\Phi}_0}\bar{S}_0(\psi_f, \psi_i)$, where $\bar{S}_0(\psi_f, \psi_i)$ is independent of $\bar{\Phi}_0$, because of 
(\ref{action of bos string-geometric model}). Here we have assumed $\sum_{p=1}^\infty
|\tilde{\bold{F}}_p|^2 =0$, which is satisfied in the perturbative string vacua because of Poincare duality, during the transition between the vacua. 
The leading term of the path-integral is given by 
\begin{equation}
\bar{Z}(\psi_f, \psi_i)=e^{\frac{i}{g_s^2}\bar{S}_0(\psi_f, \psi_i)},
\end{equation}
if we set $\frac{1}{\beta}e^{-2 \bar{\Phi}_0} \equiv \frac{1}{g_s^2}$ by renormalizing $\beta$, because $e^{-2 \bar{\Phi}_0}$ is proportional to $\frac{1}{g_s^2}$ in string theory.
If $\psi_i$ and $\psi_f$ are set minima of a potential,  
$\bar{Z}(\psi_f, \psi_i)$ becomes a transition amplitude representing a tunneling process between the semi-stable vacua with a non-perturbative correction in string coupling with the order $e^{-\frac{1}{g_s^2}}$, because $\bar{S}_0(\psi_f, \psi_i)$ becomes complex-valued in the ``classically'' forbidden region of the potential.

 \subsection{Perturbation}
 
Even in the ''tree level,'' the path-integral of string geometry theory includes higher order perturbations in string theory because string geometry includes information of genera of the world-sheets. However, string geometry theory is not only a perturbation theory, because the geometry includes all the genera and one configuration of the fields does not specify the number of genera.

We can define perturbations if we fix a ``classical'' background in the path-integral (\ref{pathint}).
The authors in \cite{Nagasaki:2023fnz, futurehetero, Kudo} identified the perturbative string vacua  in string geometry theory:
\begin{subequations}\label{eq:sec3_condition2}
\begin{align}
\bar G_{dd} &= e^{2\phi[G,B,\Phi;X]}, \\
\bar G_{d(\mu\bar\sigma)} &= 0,\\
\bar G_{(\mu\bar\sigma)(\mu'\bar\sigma')}
&= G_{(\mu\bar\sigma)(\mu'\bar\sigma')}
= \frac{\bar e^3}{\sqrt{\bar h}}
G_{\mu\nu}(X(\bar\sigma))
\delta_{\bar\sigma\bar\sigma'},\\
\bar B_{d(\mu\bar\sigma)} &= 0,\\
\bar B_{(\mu\bar\sigma)(\mu'\bar\sigma')} 
&= B_{(\mu\bar\sigma)(\mu'\bar\sigma')} 
= \frac{\bar e^3}{\sqrt{\bar h}}\,
B_{\mu\nu}(X(\bar\sigma))
\delta_{\bar\sigma\bar\sigma'},\\
\bar\varPhi &= \varPhi
= \int d \bar\sigma\hat e\Phi(X(\bar\sigma)),
\end{align}
 \end{subequations}
 where $G_{\mu\nu}(x)$, $B_{\mu\nu}(x)$ and $\Phi(x)$ represent the ten-dimensional string backgrounds, and $\phi$ satisfies a differential equation,
 \begin{eqnarray}
&- R + \frac12 |H|^2
 - 3\int d\bar\sigma\bar e
  \nabla^{(\mu\bar\sigma)}\nabla_{(\mu\bar\sigma)}\Phi
 + 3\int d\bar\sigma\bar e
  \partial_{(\mu\bar\sigma)}\Phi\partial^{(\mu\bar\sigma)}\Phi
 - \frac12\int d\bar\sigma\bar e
 \nabla^{(\mu\bar\sigma)}\nabla_{(\mu\bar\sigma)}\phi\nonumber\\
&\quad
- \frac{1}{4}\int d\bar\sigma\bar e
  \partial_{(\mu\bar\sigma)}\phi\partial^{(\mu\bar\sigma)}\phi
- 7\int d\bar\sigma\bar e
 \partial^{(\mu\bar\sigma)}\Phi\partial_{(\mu\bar\sigma)}\phi
 \nonumber\\
&
 =
 \epsilon \Biggl( \int d\bar\sigma\frac{\sqrt{\bar h}}{\bar e^2}G_{\mu\nu}
  (-\partial_{\bar\sigma}X^\mu\partial_{\bar\sigma}X^\nu
  - \bar{n}^{\bar{\sigma}}\partial_{\bar\sigma}X^{\rho}B_{\rho\mu} \bar{n}^{\bar{\sigma}} \partial_{\bar\sigma}X^{\rho'} B_{\rho'\nu}
  + \alpha'\bar e^2R_{\bar h}\Phi) 
 - i \int d\bar\sigma\frac{\sqrt{\bar h}}{\bar e^2} \bar{n}^{\bar{\sigma}} \partial_{\bar\sigma}X^\mu\nabla^\nu B_{\mu\nu} \Biggr). \nonumber\\
&
 \label{eq:sec3_def_lagrangian_L1_L2}
\end{eqnarray}
Here, we have displayed formulas in the bosonic closed sector for simplicity.
We fix a ``classical'' background to one of these perturbative vacua. By considering fluctuations around the background like $\bold{G}_{IJ} = \bar{\bold{G}}_{IJ}+ \bold{h}_{IJ}$, a Fock vacuum $|\bar{G}, \bar{B}, \bar{\Phi}, \cdots>$ is defined. 
In correlation functions, which is normalized by the path-integral, the contributions of the ``classical'' action (the zero-th order terms) are cancelled between the denominator and the numerator because the ``classical'' background is fixed and not path-integrated.    The first order terms vanish if the background is on-shell.
Thus, the action starts from the second order terms and correlation functions start from two-point correlation functions. 

In \cite{Nagasaki:2023fnz, futurehetero, Kudo},
it is shown that
two-point correlation functions of the states 
$\psi_{dd}(X, h) |\bar{G}, \bar{B}, \bar{\Phi}, \cdots>$
where 
$\psi_{MN}=\bold{h}_{MN}
-\frac{1}{2}\bar{\bold{G}}^{IJ}
\bold{h}_{IJ}\bar{\bold{G}}_{MN}$
give
the path-integrals of all-order perturbative strings on the backgrounds $\bar{G}, \bar{B}, \bar{\Phi}, \cdots$.
Namely, $\psi_{dd}$ is a state of strings. 
Scattering amplitudes of strings are given by two-point correlation functions of $\psi_{dd}(X, h) |\bar{G}, \bar{B}, \bar{\Phi}, \cdots>$ with creation operators of excited states in the first quantization.  
The path-integral over the world-sheets is generated by the summation over the world-sheets on the incoming states and the outgoing states in the two-point correlation functions because of the requirement of diffeomorphism invariance of observable.
The path-integral over the embedding functions $X^{\mu}$ is generated by insertions of the completeness relation of $X^{\mu}$ when we evaluate the time evolution in the two-point correlation functions  in the first quantization formalism.
Therefore, the path-integral of perturbative strings 
is not a priori path-integral of the string geometry theory, but a posteriori notion.

\subsection{Non-renormalization theorem}

In string geometry theory, the action does not include the derivative with respect to the coordinate $\bar{\bold{E}}_T$, including its component, two-dimensional gravitino coordinate $\bar{\chi}$ because 
$\bar{\bold{E}}_T$ is a discrete variable as in (\ref{HeteroNeighbour}). Thus, the propagator of an arbitrary fluctuation $\psi$, 
\begin{equation}
\Delta_F(\bar{\bold{E}}_T, \bold{X}_T(\bar{\tau}), \bar{\tau}; \; \bar{\bold{E}}_T,'  \bold{X}'_T(\bar{\tau}'), \bar{\tau}')=<\psi (\bar{\bold{E}}_T, \bold{X}_T(\bar{\tau}), \bar{\tau})
\psi(\bar{\bold{E}}_T,'  \bold{X}'_T(\bar{\tau}'), \bar{\tau}')>\,,
\end{equation}
satisfies
\begin{eqnarray}
&&H\left(
\frac{\pd}{\pd \bold{X}_T(\bar{\tau})}\,,
\frac{\pd}{\pd \bar{\tau}} \,;
\bar{\bold{E}}_T\,, \bold{X}_T(\bar{\tau})\,, \bar{\tau} \right)
\Delta_F(\bar{\bold{E}}_T, \bold{X}_T(\bar{\tau}), \bar{\tau}; \; \bar{\bold{E}}_T,'  \bold{X}_T(\bar{\tau}'), \bar{\tau}') \nn \\
&=&
\delta(\bar{\bold{E}}_T-\bar{\bold{E}}'_T) \delta(\bold{X}_T(\bar{\tau})-\bold{X}'_T(\bar{\tau}'))
\delta(\bar{\tau}-\bar{\tau}'), \label{delta}
\end{eqnarray}
where $H$ is the coefficient of $\psi^2$ in the action. Then, the propagator is proportional to 
$\delta(\bar{\bold{E}}_T-\bar{\bold{E}}'_T)
=
\delta(\bar{h}-\bar{h}')\delta(\bar{\chi}-\bar{\chi}')$, 
where $\bar{h}$ is the two-dimensional graviton coordinate. As a result, all the loop diagrams are proportional to $\delta(\bar{\chi}-\bar{\chi})=0$ and thus, there is no loop correction in string geometry theory. Because of no loop divergence and thus no need to renormalize, there is no problem of non-renormalizability in the path-integral of the metric on string geometry. 

\begin{figure}[htb]
\centering
\includegraphics[width=12cm]{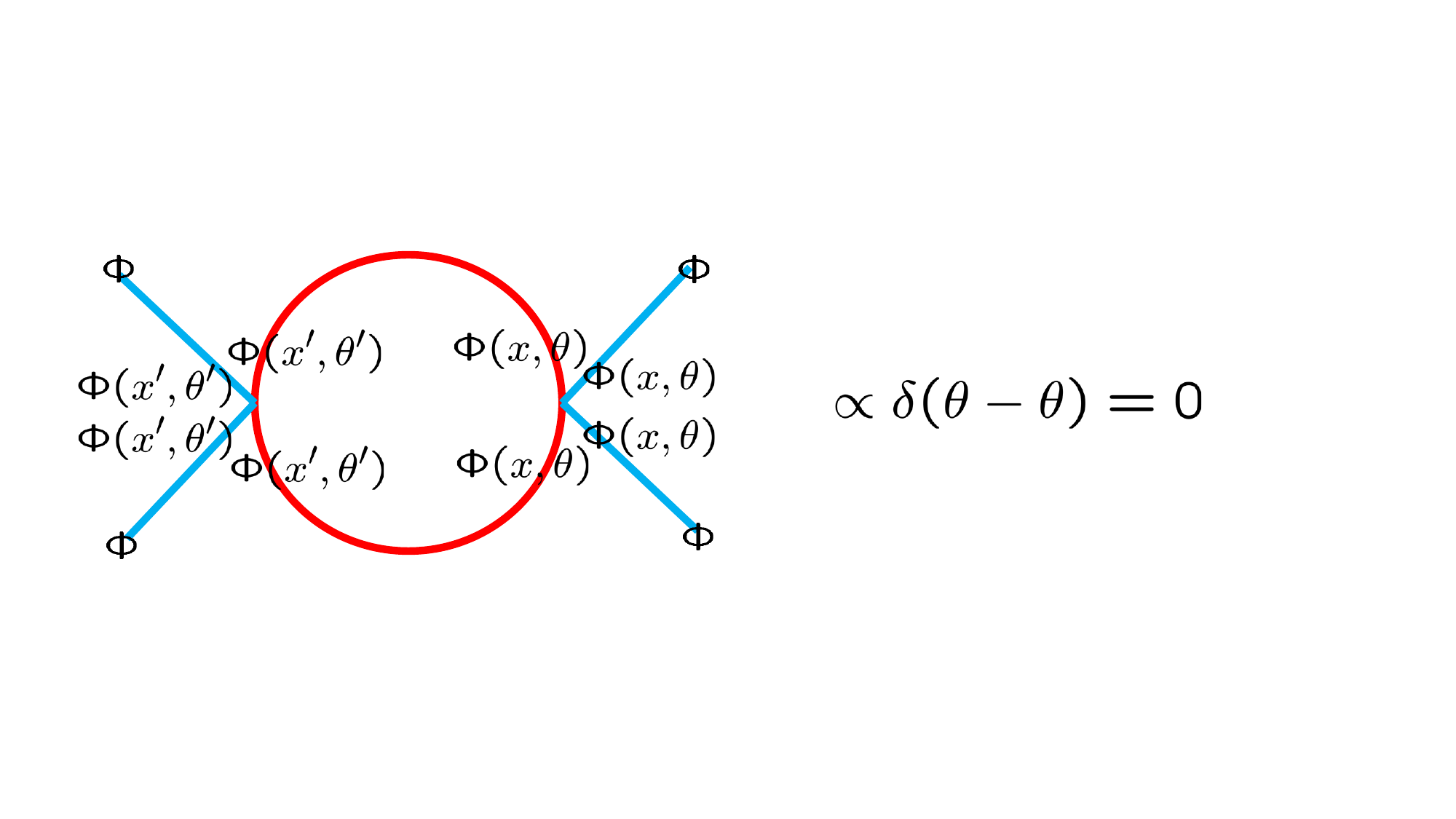}
\caption{A super loop diagram} 
\label{superdiagram}
\end{figure}
\begin{figure}[htb]
\centering
\includegraphics[width=12cm]{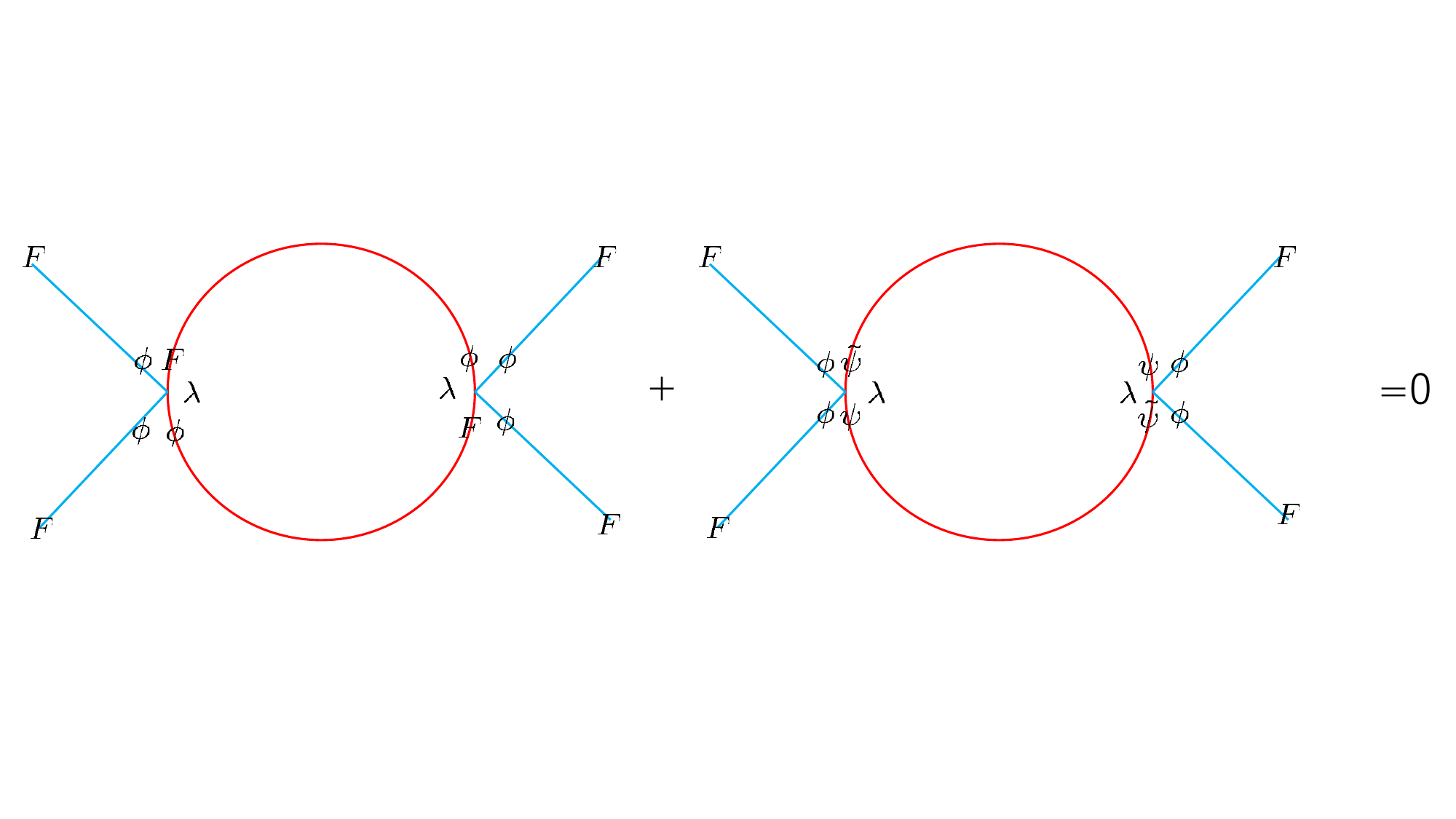}
\caption{Cancellation between  the bosonic and fermionic loop diagrams} 
\label{diagrams}
\end{figure}

Let us calculate diagrams explicitly in a simple model that has this mechanism,\footnote{Although there is no super differential term in a chiral superfield theory, the propagator between $\Phi$ and $\bar{\Phi}$ is not proportional to a delta function of Grassmann coordintes   because the chiral conditions make $\Phi$ and $\bar{\Phi}$ depend only on $\theta$ and $\bar{\theta}$, respectively. Thus, the loop diagrams that consist of only $\Phi$ are zero, whereas the loop diagrams that consist of  $\Phi$ and  $\bar{\Phi}$ are not zero.}
\begin{eqnarray}
S
&=&
\int d^p x d^2 \theta 
(\frac{1}{2}\Phi^2+\frac{1}{4!}\lambda \Phi^4)
\nonumber \\
&=&
\int d^p x
(\phi F+\psi \tilde{\psi}+\frac{1}{12}\lambda(\phi^3 F+\phi^2 \psi \tilde{\psi})
), \label{writtendown}
\end{eqnarray}
where
\begin{equation}
\Phi(x, \theta) 
=
\phi(x)+i \theta \psi(x)+ i \bar{\theta} \tilde{\psi}(x)+ \theta \bar{\theta} F(x).
\end{equation}
In Fig. \ref{superdiagram}, the super loop diagram is proportional to 
$\int d\theta \int d\theta'  \delta(\theta-\theta') \delta(\theta-\theta')=\int d\theta \delta(\theta-\theta)=0$.
In a component representation of it in Fig. \ref{diagrams}, the bosonic and fermionc loops cancel with each other, and the total loop diagrams are zero. 

 \subsection{Potentials that will determine the true vacuum in string theory}
 
The effective action of string geometry theory is the ``classical'' action itself because there is no loop correction as in subsection 3.4. Thus, only one has to do is to consider the ``classical'' potential, which does not depend on string geometry time $\bar{\tau}$, in order to determine the true vacuum. A semi-stable vacuum in the potential transits to another semi-stable vacuum with lower energy by the tunneling effect in subsection 3.2. As a result, the true vacuum in string theory will be determined by the minimum of the ``classical'' potential.  

 From the fluctuations around the perturbative vacua (\ref{eq:sec3_condition2}), the path-integrals of all order perturbative strings on the string backgrounds are derived as in subsection 3.3.  One can compare these perturbative vacua by using the ``classical'' potential.
  The authors in  
 \cite{Nagasaki:2023fnz, futurehetero, Kudo}
 restricted the classical potential in string geometry theory to the perturbative vacua and call it the potential for string backgrounds. In the particle limit $X^{(\mu \bar{\sigma})} \to x^{\mu}$ in the bosinic closed string sector, the restricted potential is given by  
\begin{align}
V_{\rm particle}
=&\frac1{2 \kappa_{10}^2}\int d^{10}x
\sqrt{-G} \nonumber \\
 & \Bigg[e^{-2\Phi+\phi}\Big(-R 
  + \frac12 |H|^2 + 2\nabla^2\phi
  + 2\partial_{\mu}\phi\partial^{\mu}\phi
  - 4\partial_{\mu}\Phi\partial^{\mu}\Phi\Big)\nonumber\\
&
 + e^{-\Phi+\frac12\phi}\big(\nabla^2\phi + \partial_{\mu}\phi\partial^{\mu}\phi\big)f \nonumber\\
&+P \left(-R
  + \frac{1}{2}|H|^2
  - 3\nabla^2 \Phi
  + 3 \partial_\mu \Phi \partial^\mu \Phi
  - \frac{1}{2} \nabla^2 \phi
  - \frac{1}{4} \partial_\mu \phi \partial^\mu \phi
  - 7 \partial^{\mu} \Phi \partial_{\mu}\phi\right) \nonumber \\
& +Q \Bigg(\nabla^2 f + (-R
  + \frac{1}{2}|H|^2
  - 3\nabla^2 \Phi
  + 3 \partial_\mu \Phi \partial^\mu \Phi
  - \frac{1}{2} \nabla^2 \phi
  - \frac{1}{4} \partial_\mu \phi \partial^\mu \phi
  - 7 \partial^{\mu} \Phi \partial_{\mu}\phi) f  \nonumber \\
& \qquad  -e^{-\Phi+\frac12\phi}
  \Big(\nabla^2 \phi 
   +\partial_{\mu}\phi\partial^{\mu}\phi\Big)\Bigg)
 \Bigg], \label{CompleteBosonicPotential}
\end{align}
where
$P$ and $Q$ are auxiliary fields. The constraints obtained by varying the potential with respect to $P$ and $Q$ are the particle limit of (\ref{eq:sec3_def_lagrangian_L1_L2}) and a formula determining $f$, respectively.  
 In these perturbative vacua, the string backgrounds need to be solutions to the equations of motion of the supergravities with stringy corrections because of  the consistency of the fluctuations, namely Weyl invariance. That is, the true vacuum will be determined by the minimum of the potential for string backgrounds where these equations of motion are imposed by the method of Lagrange multipliers.  
  The minimum must satisfy the differential equations obtained by varying the potential, which include the equations of motion of the supergravities.  If the minimum of the potential is determined among the solutions to these differential equations, the boundary condition will be determined uniquely. That is, the initial condition of the Universe will be determined by this minimum, where we can judge whether the no boundary condition \cite{Vilenkin:1982de, Hartle:1983ai, Feldbrugge:2017kzv} is correct or not for example. The string backgrounds, which parametrize the perturbative vacua, have the Lorentzian signature as they should do because only the direction of the zero mode  $x^0$ in the string coordinates $X^{(0, \sigma, \theta)}$ has the minus sign. Thus, the Universe with the time evolution will be selected as the true vacuum.

 \subsection{Time evolution}
 A perturbative string theory is given by the fluctuations around a fixed perturbative vacuum as in section 3.3. In this perturbative level,  string geometry time  $\bar{\tau}$ becomes world-sheet time   $\bar{\tau}$ and can be gauge fixed to our time $x^0$.

 In the non-perturbative level,  the Hamiltonian (\ref{hamiltonian}) is zero and thus there is no $\bar{\tau}$ time evolution, because of the constraints (\ref{Hamiltonian}), (\ref{momentum}) and (\ref{Gauss}) as in section 3.2.  

To summarize,  the time in string geometry theory is $\bar{\tau}$. In the beginning of the universe, $\bar{\tau}$ and $x^0$ are independent,
there is no $\bar{\tau}$ time evolution, instanton effects with respect to $\bar{\tau}$ cause vacuum transitions as in section 3.2, and the true vacuum will be realized. After that, $\bar{\tau}$ time evolution is non-trivial around the true vacuum, $\bar{\tau}$ and $x^0$  can be identified, and thus our  $x^0$ time evolution will be achieved.

\section{Conclusion and Discussion}

In this paper, we have solved all the problems raised in Introduction and obtained the following results. First, we have considered general fields on string geometry. We have restricted the fields to the minimal set of the fields such that their backgrounds include string backgrounds because it is natural that the backgrounds of the fields include string backgrounds in a non-perturbative formulation of string theory. The classical action is determined almost uniquely as (\ref{action of bos string-geometric model}) by T-symmetry, whose  transformation is a generalization of the T-duality transformation among 
the string backgrounds. 

Next, we have proved the non-renormalization theorem stating that there is no loop correction in string geometry theory. Thus, there is no problem of non-renormalizability, although string geometry theory is defined by the path-integral of the fields including a metric. No loop correction also results that the complete path-integrals of the all-order perturbative strings are derived from the ``tree''-level two-point correlation functions in the perturbative vacua. There is also no ``quantum'' correction to the ``classical'' potential.

Furthermore, a non-perturbative correction in string coupling with the order $e^{-\frac{1}{g_s^2}}$ is given by a transition amplitude representing a tunneling process between the semi-stable vacua  in the potential by an instanton.
Although there is no loop correction, instantons make a semi-stable vacuum transit to another semi-stable vacuum with lower energy and then, a generic state will reach the minimum of the ``classical'' potential in the end.  
 
Therefore, the conjecture that the classical potential restricted to the perturbative vacua in string geometry theory represent the string theory landscape and the minimum of the potentials gives the true vacuum in string theory, given in \cite{Nagasaki:2023fnz} is reasonable. 
Thus, next step is to search for the global minimum of the potentials. That is, we will determine an internal geometry and fluxes. One of the best analytic methods is to assume particular Calabi-Yau manifolds and flux compactifications, and then find the minimum in such a restricted region \cite{Masuda}. As a first step, the authors in \cite{Takeuchi} study a region of simple string phenomenological models and show that the minimum of our potential in this region has consistent phenomenological properties. This fact supports that our conjecture is correct. One of the best general methods is to  discretize the potential by the Regge calculus, and then find the minimum numerically \cite{Tanaka}. The fluctuations around the determined true vacuum are expected to give the Standard Model in the four dimensions plus its corrections and an inflation in the early Universe.

\section*{Acknowledgements}
We would like to thank 
D. Kadoh,
H. Kawai,
R. Kudo
T. Kugo,
H. Kunitomo
T. Masuda,
J. Nishimura
Y. Okawa,
Y. Sakatani,
S. Sasaki,
S. Sugimoto,
M. Takeuchi,
G. Tanaka, 
T. Yoneya,
and especially 
A. Tsuchiya
for discussions.

\end{document}